\documentclass[%
 reprint,
 amsmath,amssymb,
 aps,
]{revtex4-2}

\usepackage{graphicx}
\usepackage{dcolumn}
\usepackage{bm}
\usepackage{amsmath}
\allowdisplaybreaks[1]

\begin{document}

\preprint{APS/123-QED}

\title{Phase response approach to networked neural activity models with delays}

\author{Martin R Nelson}%
\email{martin.nelson@ntu.ac.uk}
\author{Jonathan J Crofts}%
\email{jonathan.crofts@ntu.ac.uk}
\affiliation{
 Department of Physical Sciences, Nottingham Trent University, UK, NG11 8NS\\
}

\date{\today}

\begin{abstract}
Networks of delay-coupled neural mass models alternate between synchrony and anti-phase locking as the delay between brain regions increases. The origin of these delay-dependent bands has remained unclear. Here, we show that a single function, the phase interaction function $H$ of an isolated node, predicts where the bands occur and explains why they repeat. For a Wilson--Cowan model with intra-nodal delay $\tau$ and inter-nodal delay $\rho$, we compute the limit cycle and phase response, and hence $H$, by harmonic balance. Each band boundary is a zero of $H'$ shifted by the phase lag $\Omega\rho$, so the bands repeat in $\rho$ with the oscillation period. Simulations of the full delay-differential networks confirm these predictions, from two nodes, where predicted frequencies agree to second order in the coupling strength, to rings, where the theory also identifies coexisting stable twisted states that random initial conditions rarely reach. At weak coupling, the delays at which synchrony gains or loses stability are set by the node dynamics alone, whereas stable anti-phase states require a bipartite network. For a macaque connectome, metastability peaks at the edges of the synchrony bands, where the stability margin of phase-locking vanishes. The reduction is accurate for coupling strengths well below the slowest Floquet exponent of the node, beyond which a phase--amplitude description is needed. More broadly, vanishing phase-locking stability provides a simple route to metastability in delay-coupled oscillator networks and a candidate mechanism for the flexible coordination of brain dynamics.
\end{abstract}

\maketitle
\section{Introduction}
Large-scale models of brain activity describe each cortical region by a neural mass model and couple the regions through an anatomical connectome, an approach implemented, for example, in The Virtual Brain~\cite{breakspear2017dynamic,sanzleon2013virtual}. Two kinds of delay arise naturally in such models: delays within a region, reflecting synaptic and dendritic processing, and delays between regions, reflecting conduction along long-range axons. In these networks, the interplay of coupling strength, delays and noise organises resting-state fluctuations and metastability~\cite{ghosh2008noise,deco2009key}, and the distribution of transmission delays determines which frequencies and phase relations emerge~\cite{petkoski2016heterogeneity,petkoski2019transmission}.

Networks of delay-coupled Wilson--Cowan nodes~\cite{WILSON19721} are a widely used example of such models (see~\cite{crofts2026review} for an introduction to the delayed model and its dynamics), and both analytical and numerical studies have examined how network structure and delays shape their dynamics~\cite{conti2019role,pinder2023PARS,pinder2024network}. A recurring finding in such networks is that the collective state changes with the delay: as the inter-regional delay is varied, the network can switch between synchronous, anti-phase and clustered states, often with several states stable at once~\cite{wang2015phase,campbell2018phase,pinder2023PARS,pinder2024network}. Such studies, whether by simulation, numerical continuation or, more recently, whole-network harmonic balance~\cite{coombes2025synchrony}, must be repeated for each network of interest. They map out where the collective state changes but do not readily explain why, and in particular which features are inherited from the dynamics of a single node and which arise from the network structure.

Weakly coupled oscillator theory provides a natural framework for separating these two contributions~\cite{kuramoto1984chemical,ashwin2016mathematical}. If each node oscillates on a stable limit cycle and the coupling is weak, the network reduces to a system of phase equations whose interactions are described by a single phase interaction function $H$, built from the infinitesimal phase response curve (iPRC) of an isolated node. Inter-nodal delays then enter only as phase lags in the argument of $H$~\cite{izhikevich1998phase,ermentrout2009delays,wang2015phase}, and the stability of synchronous and other phase-locked states follows from the sign of $H'$~\cite{earl2003synchronization}. The intra-nodal delay requires more care, since an isolated node is itself a delay-differential equation. Its iPRC is the periodic solution of a delayed adjoint equation~\cite{novivcenko2012phase,kotani2012adjoint}, which can be computed accurately by harmonic balance~\cite{nicks2024phase} or estimated by direct perturbation~\cite{winkler2022phase}.

In this paper, we apply this approach to a Wilson--Cowan network with an intra-nodal delay $\tau$ and an inter-nodal delay $\rho$, and use it to explain the behaviour found numerically in our earlier work~\cite{pinder2020oscillations,pinder2023PARS,pinder2024network}. There, at weak coupling, a ring of nodes alternated between synchronous and anti-phase behaviour in bands as $\rho$ increased, separated by narrow borders at which torus bifurcations led to weak chaos~\cite{pinder2023PARS}; a similar banded structure appeared across several network topologies and in a Macaque connectome, with high metastability near the borders~\cite{pinder2024network}. We show that these bands are set by the two zeros of $H'$, shifted by the delay, so that the pattern repeats with the oscillation period in $\rho$. At weak coupling, the delays at which synchrony gains or loses stability are the same for every network considered, whereas stable anti-phase states require a bipartite network. The theory also predicts that several twisted states (defined below) coexist stably on ring networks, and it places the high metastability of the Macaque network at the edges of the delay bands where synchrony changes stability. We test each prediction against the full delay-differential network and identify where the reduction breaks down.

The remainder of the paper is organised as follows. Section~\ref{sec:methods} introduces the delayed Wilson--Cowan network, derives its phase reduction and the interaction function $H$, and analyses the stability of phase-locked states. Section~\ref{sec:results} compares these predictions with simulations of the full network, from two nodes to rings, toy networks and a macaque connectome, and maps the limits of the weak-coupling regime. Section~\ref{sec:discussion} discusses the results and possible extensions.

\section{Model and methods}\label{sec:methods}
\subsection{Delayed Wilson--Cowan network model}\label{sec:model}
In this work, we deploy the population model due to Wilson and Cowan~\cite{WILSON19721}. The model considers two populations of excitatory and inhibitory neurons and neural activity is described by the following system
\begin{align}
    \begin{split}
     \frac{\mathrm{d}u_i}{\mathrm{d}t} &= -u_i + c_1f(u_i(t-\tau))+c_2f(v_i(t-\tau))\\&+\epsilon \sum_j w_{ij}f(u_j(t-\rho))+P,\\
     \frac{\mathrm{d}v_i}{\mathrm{d}t} &= -v_i + c_3f(u_i(t-\tau))+c_4f(v_i(t-\tau))+Q,\label{eqn:WCNetwork}
    \end{split}
\end{align}
for $i=1, 2,\ldots, n$. Here, $u_i$ and $v_i$ represent synaptic activity of the excitatory and inhibitory populations, respectively, $f$ denotes a firing rate function given by
\[
f(x) = \frac{1}{1+e^{-\beta x}},
\]
and $\epsilon$ is the coupling strength. The parameters $c_1, c_2, c_3$ and $c_4$ denote the strengths of interactions between sub-populations within a node, while $P$ and $Q$ are control parameters representing basal inputs to each node.

The coupling matrix is $W=D^{-1}A$, where $A$ is the adjacency matrix of the network ($A_{ij}=1$ if nodes $i$ and $j$ are connected and zero otherwise) and $D=\mathrm{diag}(k_1,\ldots,k_n)$ contains the node degrees. Each row of $W$ therefore sums to one, $\sum_jw_{ij}=1$, which ensures that synchronous solutions of (\ref{eqn:WCNetwork}) exist~\cite{arenas2008synchronization,nicola2021normalized,aldarabsah2021impact}. Equation~(\ref{eqn:WCNetwork}) is the voltage-based form of the model; previous studies of this network~\cite{pinder2023PARS,pinder2024network} used the equivalent activity-based form, and the change of variables relating the two is given in Appendix~\ref{app:equivalence}.

Writing $x=(u,v)^T$, a single uncoupled node obeys
\begin{equation}
\begin{split}
\frac{\mathrm{d}x}{\mathrm{d}t} &= F\big(x(t),x(t-\tau)\big)\\
&= -x(t) + C f\big(x(t-\tau)\big) + b,
\end{split}
\label{eqn:WCRHS}
\end{equation}
where
\[
C = \begin{pmatrix}c_1&c_2\\c_3&c_4\end{pmatrix},\qquad b = \begin{pmatrix}P\\Q\end{pmatrix},
\]
and $f$ acts componentwise. In the network, node $i$ receives in addition the input $\epsilon\sum_jw_{ij}f\big(u_j(t-\rho)\big)$, which is added to its excitatory equation only and does not depend on the state of node $i$ itself (since $w_{ii}=0$).

\begin{figure*}[t]
\includegraphics[width=\textwidth]{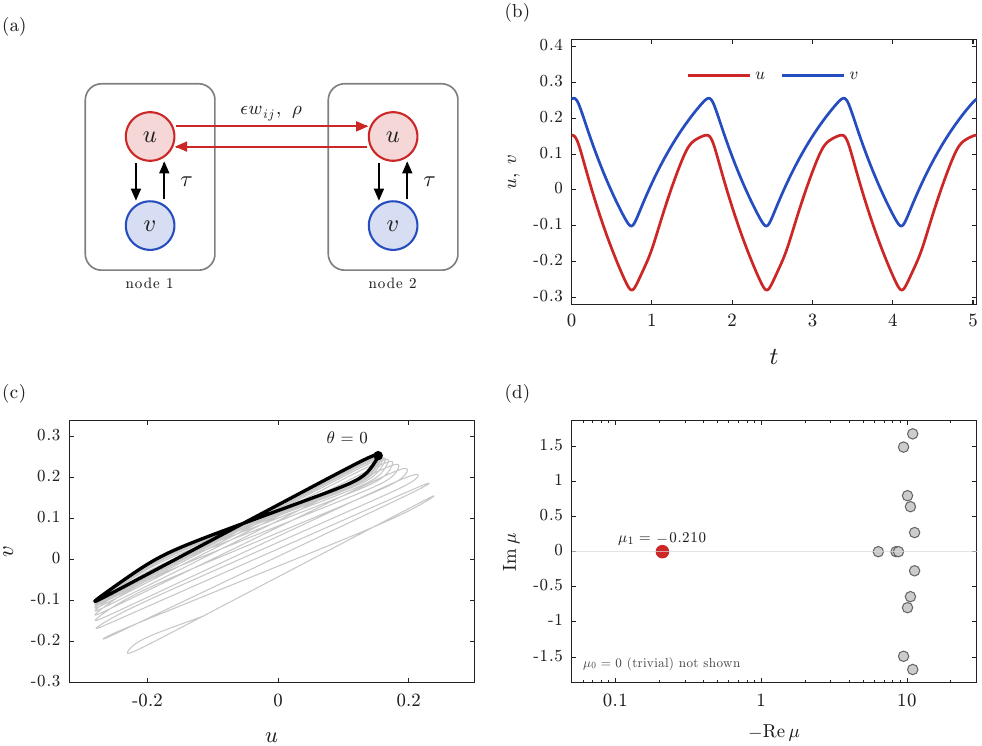}
\caption{\label{fig:singlenode}%
(a) Schematic of two delay-coupled Wilson--Cowan nodes, with intra-nodal delay $\tau$ within each node, and inter-nodal delay $\rho$ and coupling $\epsilon w_{ij}$ between excitatory populations.
(b) Time series of $u$ and $v$ on the limit cycle ($T=1.681635$) of the uncoupled node with time delay $\tau=0.5$.
(c) Phase plane showing a transient (grey) approaching the limit cycle (black); the dot marks the phase origin $\theta=0$.
(d) Converged Floquet exponents, plotted as decay rate $-\mathrm{Re}\,\mu$ (log scale) against $\mathrm{Im}\,\mu$. The slowest nontrivial exponent $\mu_1\approx-0.210$ (red) is separated from the rest of the spectrum ($-\mathrm{Re}\,\mu\gtrsim6.35$) by more than an order of magnitude; the trivial exponent $\mu_0=0$ is omitted.
Parameters: $\beta=60$, $P=0.65$, $Q=0.5$, $c_1=c_3=-1$, $c_2=-0.4$, $c_4=0$ as in \cite{pinder2023PARS,pinder2024network}.}
\end{figure*}

The above system contains two delays: an intra-node delay, $\tau$, that describes the transmission of information within each neural population, and an inter-node delay $\rho$, which represents the time taken for a signal to propagate between distant brain regions [Fig.~\ref{fig:singlenode}(a)]. Physiologically, conduction between regions takes at least as long as transmission within a region, so the biologically relevant regime is $\rho\geq\tau$. Throughout we fix $\tau=0.5$ and vary $\rho\in[0,2.5]$; the values $\rho<\tau$ are included for mathematical completeness, since none of the analysis below requires an ordering of the two delays. 

\subsection{Phase reduction of delay differential equations}\label{sec:phasereduction}
When uncoupled (\emph{i.e.} $\epsilon = 0$), each node of \eqref{eqn:WCNetwork} evolves according to \eqref{eqn:WCRHS}, which, for suitable parameter and delay values, possesses an attracting limit-cycle solution, $x_\gamma(t)$, with period $T$ and angular frequency $\omega = 2\pi/T$. For sufficiently weak coupling, the state of each node remains close to this cycle and can be described by a single phase variable $\theta_i\in[0,2\pi)$ satisfying, to first order in $\epsilon$,

\begin{equation}
    \dot{\theta}_i = \omega + \epsilon Z_u(\theta_i)\sum_{j} w_{ij}\, f\big(u_j(t-\rho)\big),
    \label{eqn:phasefull}
\end{equation}
where $Z=(Z_u,Z_v)^T$ is the infinitesimal phase response curve (iPRC) of the delayed node~\cite{novivcenko2012phase,kotani2012adjoint}; only its excitatory component $Z_u$ appears because the network input is added to the excitatory equation only. Here and below we parametrise $T$-periodic functions on the limit cycle by
the phase $\theta = \omega t \in [0, 2\pi)$, so that, for example, $Z(\theta)$ denotes $Z$ evaluated at $t = \theta/\omega$. Importantly, $x_\gamma$ and $Z$ are determined entirely by the uncoupled node: they depend on the intra-nodal delay $\tau$, but not on the coupling strength $\epsilon$ or the inter-nodal delay $\rho$.

The two delays therefore enter Eq.~(\ref{eqn:phasefull}) in different ways. The intra-nodal delay $\tau$ does not appear explicitly: its effects are contained in the frequency $\omega$ and in the shape of the iPRC $Z$, which is determined by the delayed adjoint equation (Sec.~\ref{sec:adjoint}). Because the coupling enters additively, each instant of forcing acts as an instantaneous kick to the current state, and its knock-on effect through the node's own delayed feedback is already accounted for in $Z$, which measures the asymptotic phase shift. The inter-nodal delay $\rho$ appears only through the delayed input $f\big(u_j(t-\rho)\big)$. Equation~(\ref{eqn:phasefull}) is therefore not yet closed in the phases, since its right-hand side depends on the delayed states $u_j(t-\rho)$; these are eliminated in Sec.~\ref{sec:Hfunction}, where the inter-nodal delay reduces to a constant phase lag.

\subsubsection{Adjoint equation for the intra-nodal delay}\label{sec:adjoint}
Linearising the single-node dynamics of Equation (\ref{eqn:WCRHS}) about the limit cycle, a perturbation $x_p(t)$ obeys
\begin{equation}
    \frac{\mathrm{d}x_p}{\mathrm{d}t} = DF_1(t)x_p(t) + DF_2(t)x_p(t-\tau),
\end{equation}
where $DF_1$ and $DF_2$ are the Jacobian matrices of $F$ with respect to $x(t)$ and $x(t-\tau)$, respectively, evaluated on the limit cycle~\cite{kotani2012adjoint,nicks2024phase}. For the vector field (\ref{eqn:WCRHS}) these take the simple form
\begin{equation}
\begin{split}
DF_1(t) &= -I_2,\\
DF_2(t) &= C\begin{pmatrix}
    f'\big(u_\gamma(t-\tau)\big)&0\\0&f'\big(v_\gamma(t-\tau)\big)
\end{pmatrix},
\end{split}
\end{equation}
where $I_2$ is the $2\times 2$ identity matrix. The iPRC $Z(t)$ is the $T$-periodic solution of the adjoint equation~\cite{novivcenko2012phase,kotani2012adjoint}
\begin{equation}
\begin{split}
    \dot{Z}(t) &= -DF_1^T(t)Z(t)-DF_2^T(t+\tau)Z(t+\tau)\\
    &= Z(t) - DF_2^T(t+\tau)Z(t+\tau),
\end{split}
    \label{eqn:adjoint}
\end{equation}
subject to the normalisation condition
\begin{equation}
Z^T(0)\dot{x}_\gamma(0) + \int_{-\tau}^0Z^T(s+\tau)DF_2(s+\tau)\dot{x}_\gamma(s)\,\mathrm{d}s = \omega,
\label{eqn:normalisation}
\end{equation}
which ensures that $\dot\theta=\omega$ along the unperturbed cycle (with phase measured in radians). We compute $x_\gamma$, $Z$ and the leading nontrivial Floquet exponent $\mu_1$ of the node using the harmonic balance method of Nicks \emph{et al.}~\cite{nicks2024phase}, and validate $Z$ against backward integration of (\ref{eqn:adjoint}) and the direct perturbation method~\cite{winkler2022phase}. The magnitude of $\mu_1$ quantifies the range of validity of the weak coupling approximation, which requires $\epsilon \ll |\mu_1|$.

\subsubsection{Phase interaction function and the inter-nodal delay}\label{sec:Hfunction}
Equation~(\ref{eqn:phasefull}) is not yet a closed model for the phases, since its right-hand side depends on the delayed states $u_j(t-\rho)=u_\gamma(\theta_j(t-\rho))$. Over the delay, each phase advances at a rate close to $\omega$, so $\theta_j(t-\rho)\approx\theta_j(t)-\omega\rho$. Making this replacement and averaging over one period gives the phase model~\cite{izhikevich1998phase,ermentrout2009delays,wang2015phase,campbell2018phase}
\begin{equation}
    \dot{\theta}_i = \omega + \epsilon\sum_{j=1}^n w_{ij}H(\theta_j-\theta_i-\omega\rho),
    \label{eqn:phasemodel}
\end{equation}
with phase interaction function
\begin{equation}
    H(\chi) = \frac{1}{2\pi}\int_0^{2\pi} Z_u(\theta)\,f\big(u_\gamma(\theta+\chi)\big)\,\mathrm{d}\theta.
    \label{eqn:H}
\end{equation}
The inter-nodal delay thus enters only as a phase lag.

On a phase-locked state, $\theta_i=\Omega t+\psi_i$ with constant offsets $\psi_i$, every phase advances at exactly the collective frequency $\Omega$, so the phase accumulated over the delay is exactly $\Omega\rho$~\cite{yeung1999time,earl2003synchronization}. When computing phase-locked states and their stability we therefore replace $\omega\rho$ by $\Omega\rho$ in (\ref{eqn:phasemodel}), with $\Omega$ fixed self-consistently by
\begin{equation}
    \Omega = \omega + \epsilon\sum_{j=1}^n w_{ij}H(\psi_j-\psi_i-\Omega\rho).
    \label{eqn:Omega}
\end{equation}
Phase-locked states are then described exactly, however long the delay~\cite{earl2003synchronization}; the approximation matters only for transients. Setting $\Omega=\omega$ recovers (\ref{eqn:phasemodel}), which we refer to as the limit $\epsilon\to0$. Because $\rho$ enters only through the phase lag, the stability of every phase-locked state repeats in $\rho$ with period $T$.

\subsection{Phase-locked states and their stability}\label{sec:locking}
\begin{figure*}[t]
\includegraphics[width=\textwidth]{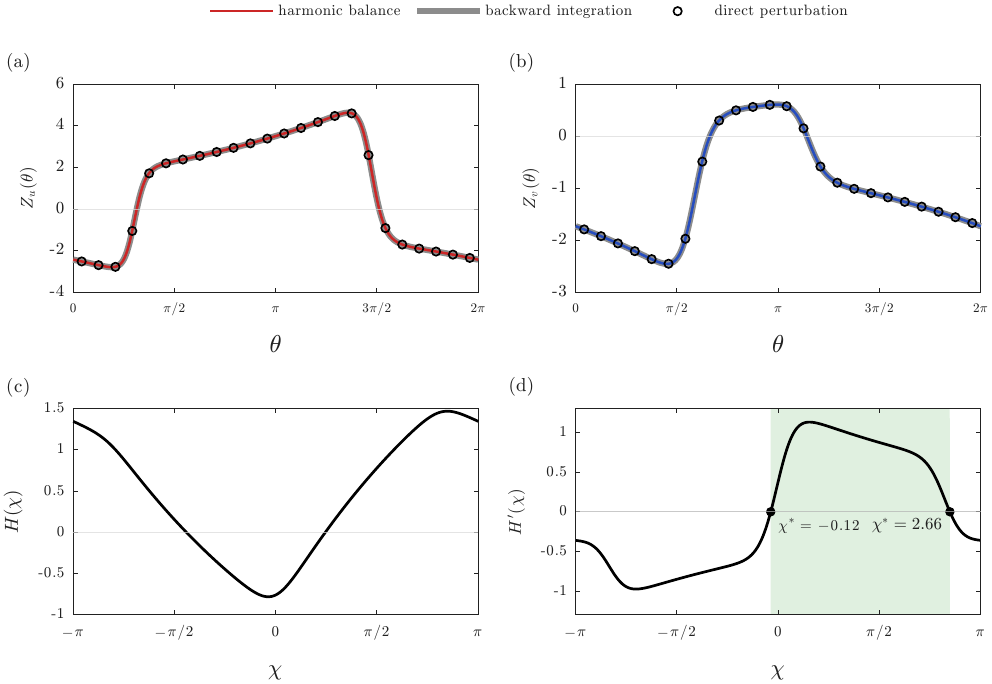}
\caption{\label{fig:prcH}%
(a), (b) Components of the infinitesimal phase response curve of the uncoupled node, computed by harmonic balance (coloured line: red for $Z_u$, blue for $Z_v$), backward integration of the adjoint equation~(\ref{eqn:adjoint}) (grey) and direct perturbation with rectangular pulses of height $0.05$ and width $0.01$ (symbols).
(c) Phase interaction function $H(\chi)$, Eq.~(\ref{eqn:H}).
(d) $H'(\chi)$; shading marks $H'>0$, with zeros $\chi^*\approx-0.12$ and $2.66$.}
\end{figure*}

\subsubsection{Linearisation: a weighted graph Laplacian}
Let $\theta_i=\Omega t+\psi_i$ be a phase-locked solution of (\ref{eqn:phasemodel}). Writing $\theta_i=\Omega t+\psi_i+\delta_i$ and linearising gives
\begin{equation}
\begin{split}
\dot{\boldsymbol\delta}&=-\epsilon\,\mathcal{L}_\psi\,\boldsymbol\delta,\\
(\mathcal{L}_\psi)_{ij}&=-w_{ij}H'(\psi_j-\psi_i-\Omega\rho)\quad(j\neq i),\\
(\mathcal{L}_\psi)_{ii}&=\textstyle\sum_{l\neq i}w_{il}H'(\psi_l-\psi_i-\Omega\rho).
\end{split}
\label{eqn:linlap}
\end{equation}
Thus, $\mathcal{L}_\psi$ is the Laplacian of the network with edge weights $w_{ij}H'(\psi_j-\psi_i-\Omega\rho)$. Its rows sum to zero, so $(1,\ldots,1)^T$ is always an eigenvector with eigenvalue $0$, reflecting invariance under a uniform phase shift; the state is linearly stable if all other eigenvalues of $\mathcal{L}_\psi$ have positive real part~\cite{chung1997spectral}.

The Laplacian of the coupling network itself, without the factors $H'$, is the normalised graph Laplacian
\begin{equation}
L=I_n-W=I_n-D^{-1}A,
\end{equation}
which is similar to the symmetric matrix 
\[I_n-D^{-1/2}AD^{-1/2}.\] 
Its eigenvalues are therefore real and, for a connected network, satisfy $0=\Lambda_1<\Lambda_2\leq\cdots\leq\Lambda_n\leq2$, with $\Lambda_n=2$ if and only if the network is bipartite. We refer to the smallest nonzero eigenvalue, $\Lambda_2$, as the spectral gap.

\subsubsection{Synchrony}
Owing to the row normalisation of $W$, (\ref{eqn:phasemodel}) admits the synchronous solution $\theta_i=\Omega t$ for all $i$, with
\begin{equation}
\Omega = \omega + \epsilon H(-\Omega\rho).
\end{equation}
Every edge then carries the same weight factor, $H'(-\Omega\rho)$, so $\mathcal{L}_\psi=H'(-\Omega\rho)\,L$ and the transverse eigenvalues of the linearisation are $-\epsilon H'(-\Omega\rho)\Lambda_k$, $k\geq2$. Since $\Lambda_k > 0$ for all $k \geq 2$ in any connected network, synchrony is stable if and only if
\begin{equation}
H'(-\Omega\rho) > 0.
\label{eqn:syncstab}
\end{equation}
When this holds, the slowest perturbations decay at rate $\epsilon H'(-\Omega\rho)\Lambda_2$. Stability of synchrony is therefore independent of topology: the network sets only the rate of approach, through $\Lambda_2$. This first-order result assumes weak coupling, identical nodes, a single inter-nodal delay and row-normalised coupling. At stronger coupling, loss of synchrony can depend on topology~\cite{pinder2024network} (Sec.~\ref{sec:limits}), and topology enters the stability condition directly if the coupling is not row-normalised or the delays differ between connections. Topology also determines which other phase-locked states coexist with synchrony, and therefore what fraction of initial conditions converge to it. The same criterion holds for directed, strongly connected networks, such as directed connectomes~\cite{crofts2022synchrony}, where the approach to synchrony may become oscillatory.

\subsubsection{Anti-phase states on bipartite networks}
If the network is bipartite, with node set $V_1\cup V_2$, then the two-cluster state $\theta_i = \Omega t$ for $i\in V_1$ and $\theta_i = \Omega t+\pi$ for $i\in V_2$ is a solution of (\ref{eqn:phasemodel}), since every neighbour of a node lies at a phase difference $\pi$; here $\Omega = \omega+\epsilon H(\pi-\Omega\rho)$. Every edge again carries the same factor, so $\mathcal{L}_\psi=H'(\pi-\Omega\rho)\,L$, and the anti-phase state is stable if and only if $H'(\pi-\Omega\rho)>0$, with the same boundaries for every bipartite network. The path, the cycle (for even $n$) and the square lattice are bipartite, whereas the complete network and the Macaque cortical network are not, and so do not support this state.

\subsubsection{Twisted states on ring networks}
On a ring of $n$ nodes each node is coupled equally to its two neighbours, so $W=\tfrac12(K+K^{T})$, where $K$ is the $n\times n$ cyclic shift matrix. Besides synchrony, the ring supports twisted states, in which the phase advances by a fixed amount $\psi$ from each node to the next,
\begin{equation}
\theta_j=\Omega t+j\psi,\qquad \psi=\frac{2\pi q}{n},\quad q=0,\ldots,n-1 .
\end{equation}
Going once round the ring must return each node to its own phase, so $\psi$ is a multiple of $2\pi/n$, and $q$ counts how many times the phase winds round the ring. Every node sees one neighbour ahead by $\psi$ and one behind by $\psi$, so all nodes receive the same input and the twisted state is a solution of (\ref{eqn:phasemodel}) with
\begin{equation}
\Omega=\omega+\frac{\epsilon}{2}\big[H(\psi-\Omega\rho)+H(-\psi-\Omega\rho)\big].
\end{equation}

Unlike synchrony and the anti-phase state, the two neighbours of a node now sit at different phase differences, so the weights in $\mathcal{L}_\psi$ depend on direction: $a=H'(\psi-\Omega\rho)$ forwards and $b=H'(-\psi-\Omega\rho)$ backwards. Then
\begin{equation}
\mathcal{L}_\psi=\tfrac12\big[a(I_n-K)+b(I_n-K^{T})\big].
\label{eqn:ringlap}
\end{equation}
Because the ring looks the same from every node, this matrix is circulant, and its eigenvectors are the Fourier modes $e^{ij\phi_k}$, $\phi_k=2\pi k/n$. Substituting $\delta_j(t)=e^{\sigma_kt}e^{ij\phi_k}$ into the linearised equation~(\ref{eqn:linlap}), with $\mathcal{L}_\psi$ given by (\ref{eqn:ringlap}), gives the growth rates
\begin{align*}
\sigma_k &= -\frac{\epsilon}{2}\Big[a\,(1-e^{i\phi_k})+b\,(1-e^{-i\phi_k})\Big],\\
\mathrm{Re}\,\sigma_k &= -\frac{\epsilon}{2}\,(a+b)\,\Lambda_k ,
\end{align*}
where $\Lambda_k=1-\cos\phi_k$ are the eigenvalues of the ring Laplacian $L$. The imaginary parts, $\tfrac\epsilon2(a-b)\sin\phi_k$, make perturbations travel round the ring but do not affect stability. The twisted state is therefore stable if and only if
\[
S_q:=H'(\psi-\Omega\rho)+H'(-\psi-\Omega\rho)>0 .
\]
The cases $q=0$ and $q=n/2$ ($n$ even), for which $a=b$, recover the synchronous and anti-phase conditions above. Twisted states are the symmetric cluster solutions of circulant networks, whose existence and stability in delayed phase models were analysed in~\cite{wang2015phase,campbell2018phase}; they are also the weak-coupling counterparts of the travelling waves found by Coombes and Meijer~\cite{coombes2025synchrony} in rings of delayed Wilson--Cowan nodes with distance-dependent coupling.

\subsection{Numerical methods}\label{sec:numerics}
The limit cycle, iPRC and slowest Floquet exponent of a node were computed by harmonic balance with $M=64$ Fourier modes~\cite{nicks2024phase}. The iPRC was checked against backward integration of the adjoint equation and against direct perturbation of the full node, and $\mu_1$ against the monodromy matrix of a Chebyshev (pseudospectral) discretisation of the delay interval~\cite{breda2012approximation}. $H$ and $H'$ were evaluated from the Fourier coefficients of $Z_u$ and $f(u_\gamma)$.

\emph{Full network.} Equations~(\ref{eqn:WCNetwork}) were integrated with MATLAB's \texttt{dde23}~\cite{shampine2001solving}, with relative tolerance $10^{-6}$ or smaller. Each node was started on the limit cycle at its chosen initial phase $\theta_j^0$, using the history $x_j(t)=x_\gamma(t+\theta_j^0/\omega)$ for $t\leq0$. An introduction to numerical methods for delayed Wilson--Cowan models, including continuation with DDE-BIFTOOL, is given in~\cite{crofts2026review}.

\emph{Phase model.} For comparison, the phase model was integrated from the same initial phases with the delayed phases retained,
\begin{equation}
\dot\theta_i=\omega+\epsilon\sum_jw_{ij}H\big(\theta_j(t-\rho)-\theta_i(t)\big),
\end{equation}
using the Euler method with step $0.01$. This form has the same phase-locked states as (\ref{eqn:phasemodel}), and the same stability conditions for synchrony and anti-phase states.

\emph{Order parameters.} The same measures are applied to both models. For the phase model the $\theta_j$ are its variables; for the full network we set $\theta_j=2\pi k$ at the $k$th maximum of $u_j$ and interpolate linearly between maxima. This defines a protophase, which coincides with the asymptotic phase at the maxima~\cite{pikovsky2001synchronization,kralemann2008phase}. Synchrony is measured by
\begin{equation}
R(t)=\Big|\frac1n\sum_je^{i\theta_j}\Big|,
\label{eqn:R}
\end{equation}
which is $1$ for synchrony and close to $0$ for a balanced anti-phase state. Anti-phase order on a bipartite network is measured by $R_{\rm bip}$, defined as in (\ref{eqn:R}) with $e^{i\theta_j}$ replaced by $s_je^{i\theta_j}$, where $s_j=\pm1$ labels the two sets. On the ring, the $q$-twisted state is detected by $R_q=|n^{-1}\sum_je^{i(\theta_j-2\pi qj/n)}|$. Finally, $\langle R\rangle$ denotes the time average of $R(t)$ over the window given in each caption, and the metastability $\langle S_R\rangle$ its standard deviation over the same window.

\section{Results}\label{sec:results}
\subsection{Single-node dynamics and validity of weak coupling}
\begin{figure*}[t]
\includegraphics[width=\textwidth]{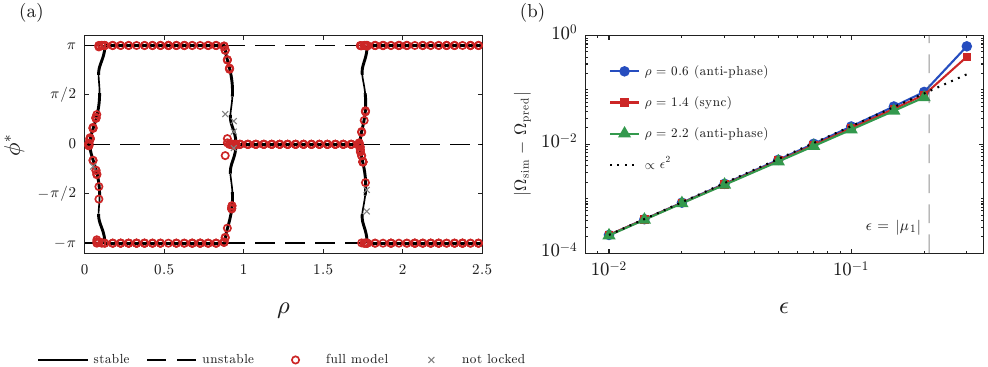}
\caption{\label{fig:twonodes}%
(a) Phase-locked states of two delay-coupled nodes at $\epsilon=0.05$ as a function of the inter-nodal delay $\rho$. Lines: phase-model prediction (solid stable, dashed unstable), using the self-consistent frequency $\Omega$. Circles: steady-state phase differences from full DDE simulations from six random initial phases per $\rho$; crosses mark runs not yet locked after $t=400$. The near-vertical branches are a mirror-image pair of non-symmetric states $\pm\phi^*$, born in pitchfork bifurcations of the synchronous and anti-phase states; in the gaps where neither synchronous nor anti-phase solutions are stable, both models lock to one of the pair. (b) Error in the collective frequency, $|\Omega_{\mathrm{sim}}-\Omega_{\mathrm{pred}}|$, against $\epsilon$ for the anti-phase ($\rho=0.6$, $2.2$) and synchronous ($\rho=1.4$) states. The dotted line is $\propto\epsilon^2$; the dashed line marks $\epsilon=|\mu_1|$.}
\end{figure*}

For the parameter values used here, which are those of~\cite{pinder2023PARS,pinder2024network}, each uncoupled node is a self-sustained oscillator (Fig.~\ref{fig:singlenode}). Harmonic balance with $M=64$ modes gives a period $T=1.681635$ (the highest retained Fourier coefficient is of order $10^{-11}$). Because the two formulations are related by the change of variables of Appendix~\ref{app:equivalence}, this is also the period of the oscillations studied in~\cite{pinder2023PARS,pinder2024network}. The Floquet spectrum, computed both from the monodromy operator and from the harmonic-balance eigenvalue problem, contains a single slow nontrivial exponent, $\mu_1=-0.2098$, separated from the rest of the spectrum ($\mathrm{Re}\,\mu\lesssim-6.35$) by more than an order of magnitude [Fig.~\ref{fig:singlenode}(d)]. Perturbations off the cycle therefore relax along essentially one direction, on a time scale $1/|\mu_1|\approx5$, which is comparable to three periods. This sets the range of validity of the phase reduction: the weak-coupling assumption requires $\epsilon\ll|\mu_1|\approx0.21$. The banded structure reported in~\cite{pinder2023PARS,pinder2024network} for $\epsilon\lesssim0.2$ therefore lies at the upper edge of this range, and we expect quantitative agreement to degrade as $\epsilon$ approaches $|\mu_1|$.

\subsection{Phase response curves and the phase interaction function}
\begin{figure*}[t]
\includegraphics[width=\textwidth]{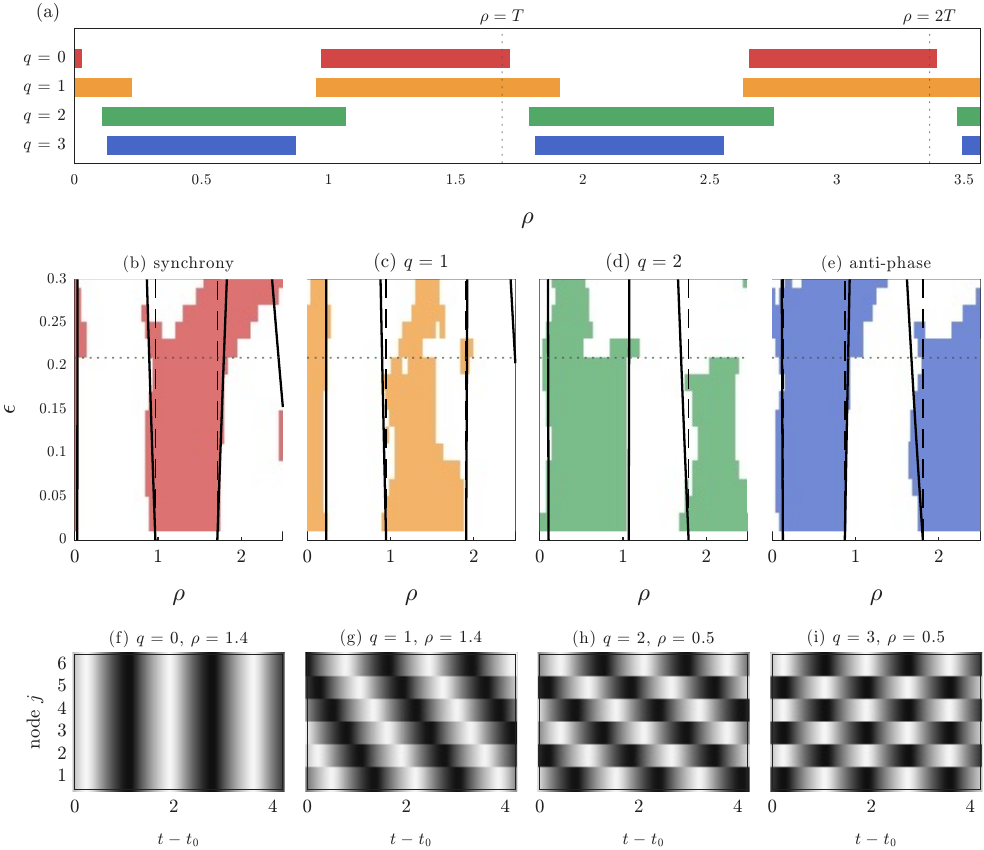}
\caption{\label{fig:ring6}%
Twisted states on a ring of $n=6$ nodes.
(a) Intervals of the inter-nodal delay $\rho$ in which each twisted state $\theta_j=\Omega t+2\pi qj/6$, $q=0,\ldots,3$, is linearly stable in the phase model as $\epsilon\to0$; $q=0$ is synchrony and $q=3$ the anti-phase state. The pattern repeats with period $T$ in $\rho$, and the intervals overlap, so multiple twisted states are stable at the same delay.
(b)--(e) Full DDE network started close to twist $q$ (initial phases $2\pi qj/6$ plus noise of standard deviation $0.1$, integrated to $t=600$; one realisation per cell): shaded cells mark parameters at which the twisted pattern persists (twist order parameter $R_q\geq0.9$ over the final $60$ time units). Solid lines: predicted stability boundaries using the self-consistent frequency $\Omega$ (including, at top right of (b) and (c), the edge of the next band); dashed lines: the limit $\epsilon\to0$ ($\Omega=\omega$); dotted line: $\epsilon=|\mu_1|$. For $\epsilon\leq0.1$ the full model agrees with the prediction at 93--96\% of grid points.
(f)--(i) Space--time plots of $u_j(t)$ (dark: high activity) for an example of each twisted state in the full model at $\epsilon=0.05$, over $2.5$ periods after a transient of $t_0\approx296$. Panels (f) and (g) share the same parameters ($\rho=1.4$), as do (h) and (i) ($\rho=0.5$); only the initial phases differ.}
\end{figure*}

The iPRC obtained by harmonic balance agrees with backward integration of the adjoint equation to a relative accuracy of $2\times10^{-7}$, and with direct perturbation to within $0.5\%$ of its peak-to-peak amplitude [Fig.~\ref{fig:prcH}(a),(b)]. The harmonic-balance solution converges spectrally: its maximum difference from a reference solution with $M=128$ modes is $8\times10^{-2}$, $1.4\times10^{-3}$ and $3\times10^{-7}$ for $M=16$, $32$ and $64$. The steep firing-rate function ($\beta=60$) makes both $Z_u$ and $Z_v$ change rapidly over a short fraction of the cycle, so moderate truncations are not sufficient. The resulting interaction function $H$ [Fig.~\ref{fig:prcH}(c)] is close to a shifted sinusoid, and $H'$ has exactly two zeros, $\chi^*\approx-0.12$ and $2.66$ [Fig.~\ref{fig:prcH}(d)]. Synchrony is stable when $-\Omega\rho$ (mod $2\pi$) lies in the shaded interval, and the anti-phase state when $\pi-\Omega\rho$ does. Every stability boundary in the remainder of the paper follows from these two numbers. With $\Omega=\omega$, synchrony is stable for $\rho\in[0,0.031)\cup(0.969,1.713)$ and the anti-phase state for $\rho\in(0.128,0.872)\cup(1.810,2.554)$, the pattern repeating with period $T$ in $\rho$. The synchronous ($\rho=1.4$) and anti-phase ($\rho=0.6$) solutions observed in~\cite{pinder2023PARS} for $\epsilon=0.1$ both lie inside the predicted bands, and the repeating banded structure reported there is explained directly by this analysis.

\subsection{Two coupled nodes}
\begin{figure*}[t]
\includegraphics[width=\textwidth]{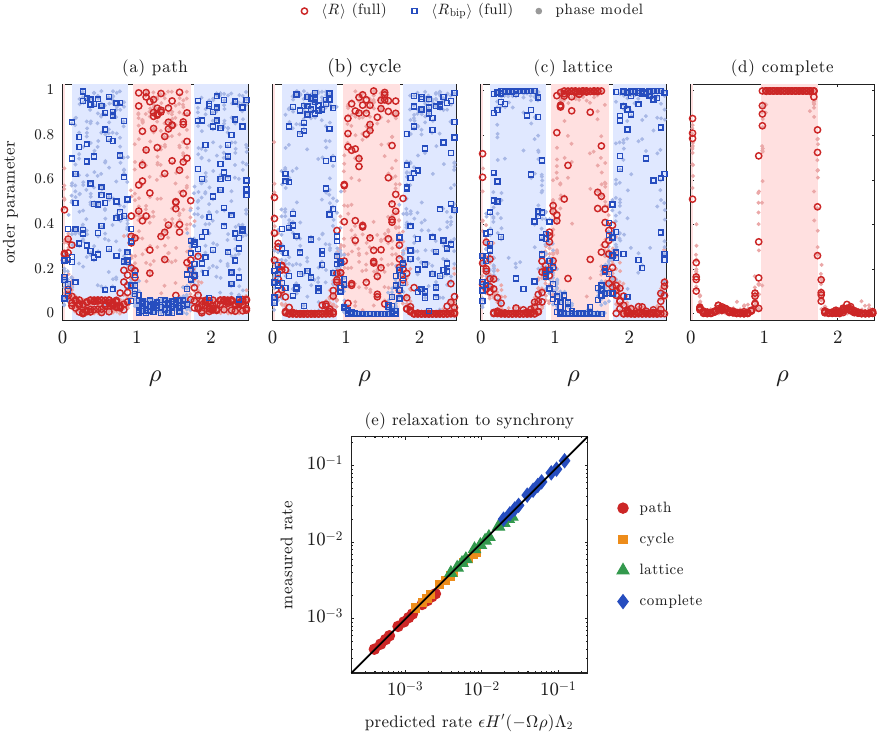}
\caption{\label{fig:toy}%
Toy networks of $n=16$ nodes at $\epsilon=0.05$.
(a)--(d) Order parameters against $\rho$ for the path, cycle, $4\times4$ lattice and complete network: $\langle R\rangle=\langle|n^{-1}\sum_j e^{i\theta_j}|\rangle$ (circles) and, for the three bipartite networks, $\langle R_{\mathrm{bip}}\rangle=\langle|n^{-1}\sum_j s_je^{i\theta_j}|\rangle$ (squares), where $s_j=\pm1$ labels the two node sets; $\langle R\rangle\approx0$ for the anti-phase state, so $R_{\mathrm{bip}}$ is needed to detect it. Phases are defined from successive maxima of $u_j$, and averages are taken over $300\leq t\leq400$ from four random initial conditions per $\rho$. Small dots: delayed phase model run from the same and additional initial phases for the same time. Shading marks predicted stability of synchrony, $H'(-\Omega\rho)>0$ (red), and of the anti-phase state, $H'(\pi-\Omega\rho)>0$ (blue); the boundaries are the same for every topology.
(e) Measured rate of relaxation to synchrony (full model, started along the eigenvector of $L$ for the spectral gap $\Lambda_2$) against the prediction $\epsilon H'(-\Omega\rho)\Lambda_2$, for $\rho\in\{1.1,1.25,1.4,1.55\}$ and $\epsilon\in\{0.025,0.05,0.1\}$; the line is $y=x$.}
\end{figure*}

For two nodes the phase difference $\phi=\theta_2-\theta_1$ obeys
\[
\dot\phi=\epsilon\big[H(-\phi-\Omega\rho)-H(\phi-\Omega\rho)\big],
\]
whose zeros are the phase-locked states. Since this is a flow on a circle, two identical nodes always phase-lock, and Fig.~\ref{fig:twonodes}(a) shows which state they select as $\rho$ varies. Synchrony ($\phi=0$) and the anti-phase state ($\phi=\pi$) are stable in alternating bands, separated by narrow windows, $\rho\approx0.03$--$0.13$, $0.88$--$0.95$ and $1.73$--$1.77$, in which neither is stable. Each window is bounded by pitchfork bifurcations: where synchrony or the anti-phase state loses stability, a stable pair of locked states $\pm\phi^*$ branches from it, sweeps across the window and merges with the other state as that state gains stability (the near-vertical branches). The two states are mirror images, in which either node leads. At every $\rho$, therefore, the nodes synchronise, lock in anti-phase, or lock onto one of this pair. Simulations of the full DDE at $\epsilon=0.05$ agree closely with this picture throughout $0<\rho<2.5$, including the windows, where the full model locks onto $+\phi^*$ or $-\phi^*$ depending on the initial condition.

Symmetry alone guarantees that synchrony and the anti-phase state exist at any $\epsilon$, since $\phi=0$ and $\pi$ are exact solutions for two identical nodes; what the phase model predicts is where they are stable and the values of $\phi^*$ in between. A further, quantitative test is the collective frequency, which the phase model predicts to be
\[
\Omega=\omega+\tfrac{\epsilon}{2}\big[H(\phi^*-\Omega\rho)+H(-\phi^*-\Omega\rho)\big].
\]
The error in this prediction scales as $\epsilon^2$ (fitted slopes $1.94$--$1.98$ for $\epsilon\leq0.05$), as expected for a first-order theory, and stays below $2\%$ of $\Omega$ up to $\epsilon=0.2$ [Fig.~\ref{fig:twonodes}(b)]. It grows abruptly at $\epsilon=0.3$, close to the Hopf bifurcation of the synchronous state~\cite{pinder2023PARS}. The $\epsilon^2$ coefficient is the same for all three states, which suggests a common origin: the mean coupling input displaces the limit cycle along the slow Floquet direction, independently of the phase dynamics.

\subsection{Ring networks: alternating synchronous and anti-phase bands}
\begin{figure*}[t]
\includegraphics[width=\textwidth]{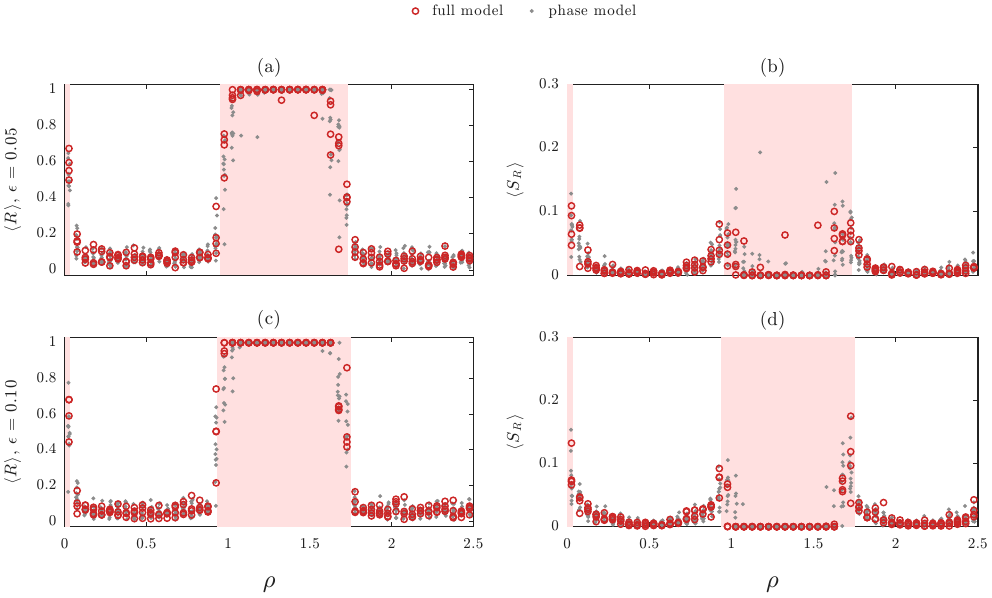}
\caption{\label{fig:macaque}%
Macaque cortical network (CoCoMac, $n=47$). (a), (c) Mean order parameter $\langle R\rangle$ and (b), (d) metastability $\langle S_R\rangle$, the standard deviation of $R(t)$ over $300\leq t\leq500$, against $\rho$ for $\epsilon=0.05$ (top) and $\epsilon=0.1$ (bottom). Circles: full DDE network from four random initial conditions per $\rho$; dots: delayed phase model from the same and additional initial phases. Shading marks predicted stability of synchrony, $H'(-\Omega\rho)>0$. In both models metastability peaks at the predicted stability boundaries.}
\end{figure*}

The phase model explains the alternating synchronous and anti-phase bands found on a six-node ring~\cite{pinder2023PARS} (Fig.~\ref{fig:ring6}). The ring has four distinct twisted states, $q=0,\dots,3$ ($q=0$ is synchrony, $q=3$ anti-phase), each stable where $S_q>0$. As $\epsilon\to0$ their stable intervals repeat with period $T$ in $\rho$ [Fig.~\ref{fig:ring6}(a)], so the band spacing in~\cite{pinder2023PARS} is the oscillation period, and the chaotic borders lie where $S_q$ changes sign. Full-network simulations seeded near each twisted state confirm these predictions at $93$--$96\%$ of grid points for $\epsilon\leq0.1$ [Fig.~\ref{fig:ring6}(b)--(e)]; the remaining points lie next to borders, where weakly stable states converge slowly.

The phase model also predicts multistability that earlier simulations did not resolve: the states $q=1$ and $q=2$ are stable over bands as wide as those of synchrony and anti-phase, overlapping them near $\rho\approx1.4$ and $\rho\approx0.5$ respectively. Figure~\ref{fig:ring6}(f)--(i) shows this in the full network: depending only on the initial condition, the ring settles into synchrony or a travelling wave with phase step $\pi/3$ [(f), (g)], or into three or two phase clusters [(h), (i)]. The twisted states have small basins, however. From random initial phases at $\epsilon\leq0.1$ ($1000$ runs), the full network reaches the anti-phase state in $41\%$ of runs and synchrony in $21\%$, but $q=2$ and $q=1$ in only $5\%$ and $3\%$; in the remaining $29\%$ the network was still oscillating but had not settled onto a twisted state by $t=300$. This is why only synchronous and anti-phase solutions were reported in~\cite{pinder2023PARS}. Above $\epsilon\approx|\mu_1|$ the stable regions of the full network widen beyond the predicted boundaries, the first sign of the breakdown examined in Sec.~\ref{sec:limits}.

\subsection{Toy networks: bipartite structure and spectral gap}
\begin{figure*}[t]
\includegraphics[width=\textwidth]{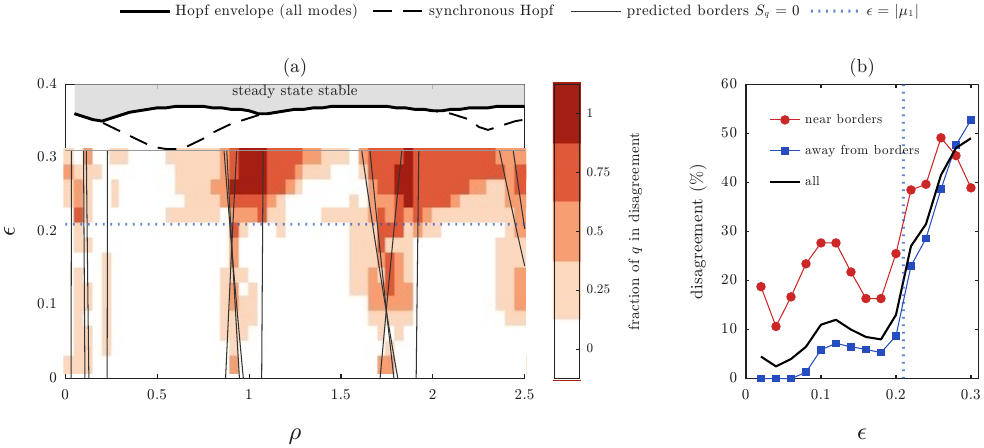}
\caption{\label{fig:validity}%
Breakdown of the phase reduction for the ring of six nodes. (a) Fraction of the twisted states $q=0,\ldots,3$ for which the persistence of a seeded state in the full model (Fig.~\ref{fig:ring6}) disagrees with the first-order prediction, over $(\rho,\epsilon)$. Thin lines: predicted stability borders $S_q=0$; dotted line: $\epsilon=|\mu_1|$; thick solid line: Hopf envelope of the network steady state over all network modes (dashed: synchronous mode only), above which oscillations are quenched (grey). Simulations extend to $\epsilon=0.3$. (b) Percentage of disagreement against $\epsilon$ for points within $0.1$ in $\rho$ of a predicted border, away from the borders, and overall.}
\end{figure*}

Pinder \emph{et al.}~\cite{pinder2024network} observed that, for $\epsilon\lesssim0.2$, the path, cycle, lattice and complete networks all show a similar central band of high synchrony, but with ``robust'' synchrony on the lattice and complete network and only ``moderate'' synchrony and metastability on the path and cycle. Section~\ref{sec:locking} explains the first observation: at weak coupling, synchrony is stable if and only if $H'(-\Omega\rho)>0$ on every network considered, so the band edges must coincide. This is confirmed in Fig.~\ref{fig:toy}(a)--(d), where the full model and the phase model change behaviour at the same predicted delays on all four networks. The path, the cycle ($n=16$) and the lattice are bipartite, so they also support the anti-phase state, stable when $H'(\pi-\Omega\rho)>0$; the complete network is not, and outside its synchronous band it shows only weakly ordered states ($\langle R\rangle\approx0.05$), consistent with the ``out-of-phase'' oscillations reported in~\cite{pinder2024network}. Because $\langle R\rangle\approx0$ for a balanced anti-phase state, the Kuramoto order parameter alone cannot distinguish it from incoherence; the bipartite order parameter $R_{\mathrm{bip}}$ shows that much of the apparently low synchrony outside the central band is in fact anti-phase locking. The second observation reflects the rate at which synchrony is approached, $\epsilon H'(-\Omega\rho)\Lambda_2$, which differs by a factor of about $50$ between the path (spectral gap $\Lambda_2=0.022$) and the complete network ($1.067$). This rate is reproduced by the full model over almost three orders of magnitude. [Fig.~\ref{fig:toy}(e); median relative error $2$--$3\%$ for $\epsilon\leq0.05$]. On the path, a finite simulation therefore samples a slow transient, and the phase model run for the same time reproduces the spread of $\langle R\rangle$; on the cycle, part of the spread instead reflects coexisting stable twisted states, as on the six-node ring. The ``moderate synchrony'' of the sparse networks is thus not a distinct dynamical regime but a consequence of slow convergence and multistability, both controlled by the network spectrum.

\subsection{Macaque cortical network}
For the Macaque connectome, Pinder \emph{et al.}~\cite{pinder2024network} noted a striking resemblance to the complete network at weak coupling. Both networks are non-bipartite (largest Laplacian eigenvalue $\Lambda_n=1.42<2$ for the connectome), so neither supports the anti-phase state, and at weak coupling the synchronous band is the same on both; the resemblance is therefore expected. Figure~\ref{fig:macaque} shows that the phase model reproduces the full model's $\langle R\rangle$ both inside and outside the predicted band at $\epsilon=0.05$ and $0.1$. It also reproduces the metastability $\langle S_R\rangle$, which in both models is small deep inside and far outside the band but peaks at its edges. In the weak-coupling regime, high metastability is thus located where the stability margin $H'(-\Omega\rho)$ of synchrony vanishes and the relaxation becomes slow, and it is predicted by the zeros of a single function of the uncoupled node. The phase-cluster dynamics and the island of high metastability found in~\cite{pinder2024network} at $(\rho,\epsilon)=(1.75,0.37)$ lie well beyond $\epsilon=|\mu_1|$, close to the Hopf envelope, and are outside the scope of the present reduction.

\subsection{Limits of the weak coupling regime}\label{sec:limits}
Figure~\ref{fig:validity} summarises the comparison of Fig.~\ref{fig:ring6} as a single measure of disagreement between the phase model and the full network over the $(\rho,\epsilon)$ plane. Two coupling strengths organise this plane. The first is $|\mu_1|\approx0.21$. Away from the predicted borders, the disagreement stays below about $7\%$ for $\epsilon\lesssim0.2$ and jumps to $40$--$50\%$ once $\epsilon$ exceeds $|\mu_1|$ [Fig.~\ref{fig:validity}(b)], so the slowest Floquet exponent is not just a formal bound but the coupling at which the reduction actually fails. Near the borders, where $S_q$ is small and the dynamics are slow, the disagreement is larger at every $\epsilon$; these are also the parameters at which torus breakdown and weak chaos were found in~\cite{pinder2023PARS,pinder2024network}. The second is the Hopf bifurcation of the network steady state, which lies at $\epsilon\approx0.35$--$0.37$ when all network modes are included (computed from the rightmost root of the linearised DDE). For the synchronous mode alone this reproduces the range $\epsilon\in[0.3125,0.3705]$ found by numerical continuation in~\cite{pinder2023PARS}, which also confirms that the two formulations of the model are equivalent. Between $|\mu_1|$ and the Hopf bifurcation lies an intermediate regime, containing the higher-codimension bifurcations, multistability and phase-cluster states reported in~\cite{pinder2023PARS,pinder2024network}, which a phase-only description cannot capture. Because the node has a single slow, real Floquet exponent, a phase--amplitude reduction retaining $\mu_1$ is the natural extension into this regime~\cite{nicks2024phase}.

\section{Discussion and Conclusions}\label{sec:discussion}
We have shown that the banded structure found numerically in delayed Wilson--Cowan networks at weak coupling~\cite{pinder2023PARS,pinder2024network} is governed by a single function of the isolated node: the phase interaction function $H$. Every stability boundary is a zero of $H'$, shifted by the phase lag $\Omega\rho$, so the bands repeat with the oscillation period in $\rho$. At weak coupling, the delays at which synchrony gains or loses stability are set by the node dynamics alone, while the network topology sets the rate of convergence and determines which other phase-locked states exist: anti-phase states require a bipartite network, and rings support twisted states that are stable over wide ranges of $\rho$ but are rarely reached from random initial conditions. Together, these results show that the ``moderate'' synchrony reported for sparse networks reflects slow convergence and multistability rather than a distinct dynamical regime, and that the high metastability of the Macaque network at weak coupling is located where the stability margin of synchrony vanishes. In each case the prediction requires computing $H$ once, rather than a sweep of the full delay-differential network.

The main limitation of the approach is that it describes each node by its phase alone. The reduction is accurate for $\epsilon$ well below the slowest Floquet exponent, $|\mu_1|\approx0.21$, but the dynamics found between this scale and the Hopf envelope, including torus breakdown, weak chaos and phase-cluster states~\cite{pinder2023PARS,pinder2024network}, involve deviations from the limit cycle that a phase model cannot represent. Because the Floquet spectrum of the node contains a single slow real exponent, well separated from the rest, a phase--amplitude reduction retaining just this one amplitude direction~\cite{nicks2024phase} is a natural next step; it should also account for the $\mathcal{O}(\epsilon^2)$ frequency shift observed for two nodes, which we attribute to the mean coupling input displacing the cycle along this direction. Alternatively, beyond weak coupling the periodic states of the full network and their Floquet stability can be computed directly by harmonic balance, which for circulant networks decouples into independent network modes~\cite{coombes2025synchrony}. The same limitation qualifies our statements about network structure. The topology independence of the synchronous band is a first-order, weak-coupling result: once amplitude effects matter, each network mode has its own Floquet multipliers, and modes with Laplacian eigenvalues close to $2$ can destabilise synchrony before the others, as the difference between the Hopf envelopes of the synchronous mode and of all ring modes already shows (Sec.~\ref{sec:limits}). This is consistent with earlier numerical observations that network structure can destabilise synchrony at stronger coupling~\cite{pinder2024network}.

Several other extensions are straightforward within the present framework. Delays that differ between connections, as expected from the distances between brain regions~\cite{petkoski2016heterogeneity,petkoski2019transmission}, enter as edge-dependent phase lags $\Omega\rho_{ij}$ with the same function $H$; exact synchrony then generally no longer exists, and the near-synchronous state and its stability depend jointly on the network and the distribution of delays. More generally, conduction delays are distributed rather than fixed. For a distribution $g(s)$ of inter-nodal delays, the node, its iPRC and $H$ are unchanged, and the phase lag is simply averaged over the distribution: $H(\chi-\Omega\rho)$ is replaced by $H_g(\chi)=\int_0^\infty g(s)\,H(\chi-\Omega s)\,\mathrm{d}s$. Writing $H(\chi)=\sum_k h_k e^{ik\chi}$, this multiplies each harmonic by $\hat g(k\Omega)$, where $\hat g(\omega)=\int_0^\infty g(s)e^{-i\omega s}\,\mathrm{d}s$. A spread of delays comparable with the period therefore suppresses the higher harmonics of $H$, shifting the band edges and narrowing the stability margins of all phase-locked states. A distribution of intra-nodal delays instead changes the node itself; Winkler \emph{et al.}~\cite{winkler2022phase} have shown how to compute the iPRC and phase interaction function of Wilson--Cowan oscillators in this case, for Gaussian and log-normal delay distributions, and their results can be inserted directly into the analysis above. Because $H$ depends on $\tau$, the band structure over the whole $(\tau,\rho)$ plane can also be mapped, at the cost of one computation of $H$ per value of $\tau$.

Delays need not be constant either: conduction velocity depends on myelination, which is itself activity dependent, so inter-regional delays are state dependent. Jolly \emph{et al.}~\cite{jolly2026phase} have recently developed a framework for phase-oscillator networks with multiple and state-dependent delays, motivated by white-matter plasticity, and found that delay plasticity can drive networks towards more coherent behaviour. Combining that framework with the interaction function computed here would give a model, derived from a neural mass model rather than postulated, in which the delays and the phase-locked states they select evolve together.

The framework also extends to directed and weighted connectomes, for which the Laplacian has complex eigenvalues. With row-normalised coupling the synchrony bands are unchanged, but relaxation becomes oscillatory, with a rate set by $\mathrm{Re}\,\Lambda_2$ of the directed Laplacian; this may account for the slower synchronisation found in directed connectomes~\cite{crofts2022synchrony}. The existence and stability of other phase-locked states, and hence the sizes of their basins of attraction, also depend on the directed structure, as found for the attractor states of Hopfield networks on directed connectomes~\cite{padmore2020modelling}.

More broadly, our results identify vanishing phase-locking stability as a simple route to metastability in delay-coupled oscillator networks: close to a zero of $H'$, perturbations decay slowly and the network lingers near states that are only weakly stable or weakly unstable. Since $H$ can be computed for any neural mass model with a stable limit cycle, including those with intra-nodal delays, the same analysis applies directly to other models used in whole-brain simulations, and it suggests a mechanism for the flexible coordination of brain dynamics that can be tested in data-driven connectome models.

\appendix
\section{Equivalence with the activity-based formulation}\label{app:equivalence}
Previous studies of this network~\cite{pinder2023PARS,pinder2024network} employed the activity-based formulation~\cite{coombes2009delays}:
\begin{equation}
\begin{split}
     \frac{\mathrm{d}\hat u_i}{\mathrm{d}t} &= -\hat u_i + f\Big(c_1\hat u_i(t-\tau)+c_2\hat v_i(t-\tau)+P\\
     &\qquad+\epsilon\sum_j w_{ij}\hat u_j(t-\rho)\Big),\\
     \frac{\mathrm{d}\hat v_i}{\mathrm{d}t} &= -\hat v_i + f\big(c_3\hat u_i(t-\tau)+c_4\hat v_i(t-\tau)+Q\big),
\end{split}
\label{eqn:WCactivity}
\end{equation}
in which the coupling acts inside the firing rate function. The two formulations are equivalent. Given a solution of (\ref{eqn:WCactivity}), define the total synaptic inputs
\begin{align*}
u_i(t) &= c_1\hat u_i(t)+c_2\hat v_i(t)+P\\
&\quad+\epsilon\sum_j w_{ij}\hat u_j(t-\rho+\tau),\\
v_i(t) &= c_3\hat u_i(t)+c_4\hat v_i(t)+Q,
\end{align*}
so that (\ref{eqn:WCactivity}) reads $\dot{\hat u}_i = -\hat u_i + f(u_i(t-\tau))$, $\dot{\hat v}_i = -\hat v_i + f(v_i(t-\tau))$. Differentiating and substituting shows directly that $(u_i,v_i)$ satisfy (\ref{eqn:WCNetwork}) with the same parameters and delays. (For $\rho<\tau$ the map involves values of $\hat u_j$ at later times, so it relates complete solutions of the two systems rather than their initial histories; this does not affect periods, phase relations or asymptotic states.) Conversely, provided $\det C = c_1c_4-c_2c_3\neq0$ (for the parameters used here $\det C = -0.4$), this linear relation can be inverted, so solutions of the two systems are in one-to-one correspondence. Since $(u_i,v_i)$ is a linear combination of the state of node $i$ up to an $\mathcal{O}(\epsilon)$ term, the period of the uncoupled oscillation and the phase relationships between nodes are preserved, and results obtained for (\ref{eqn:WCNetwork}) apply directly to the previously studied system (\ref{eqn:WCactivity}). The advantage of (\ref{eqn:WCNetwork}) for the present work is that the coupling is additive and independent of the receiving node, which simplifies the phase reduction of Sec.~\ref{sec:phasereduction}.

\begin{acknowledgments}
We thank Stephen Coombes and Rachel Nicks (University of Nottingham) for helpful discussions, and in particular for suggesting weakly coupled oscillator theory as a way to analyse the delayed Wilson--Cowan networks studied in our earlier work~\cite{pinder2023PARS,pinder2024network}.
\end{acknowledgments}

\bibliography{Crofts26_PRE_WC}

\end{document}